\documentstyle[12pt]{article}
  \advance\oddsidemargin  by -1in
  \advance\evensidemargin by -1in
  \advance\topmargin      by -0.5in
\def\lsim{\mathrel{\rlap{\lower4pt\hbox{\hskip1pt$\sim$}}
    \raise1pt\hbox{$<$}}}         
\def\gsim{\mathrel{\rlap{\lower4pt\hbox{\hskip1pt$\sim$}}
    \raise1pt\hbox{$>$}}}         

\def\beq{\begin{equation}}
\def\endeq{\end{equation}}
\def\arr{\begin{eqnarray}}
\def\endarr{\end{eqnarray}}
\phantom{.}

\makeindex
\begin{document}
\vskip1cm

\pagestyle{empty}
\hfill{\large DFTT 64/93}

\hfill{\large October 1993}

\vspace{2.0cm}

\begin{center}

\centerline{{\bf PERTURBATIVE QCD FORBIDDEN CHARMONIUM DECAYS}}
\vskip0.5cm
\centerline{{\bf AND GLUONIA}}
\vskip1cm

M.~Anselmino$^{a}$, M.~Genovese$^{a}$ and D.E.~Kharzeev$^{b,c}$
\vskip0.3cm

$^{a}$Dipartimento di
Fisica Teorica, Universit\`a di Torino

and INFN, Sezione di
Torino, 10125 Torino, Italy
\medskip

$^{b}$ INFN, Sezione di Pavia, 27100 Pavia, Italy
\medskip

$^{c}$ Moscow University,
119899, Moscow, Russia

\vspace{1.0cm}

{\large \bf Abstract \bigskip }
\end{center}

We address the problem of observed charmonium decays which should be
forbidden in perturbative QCD. We examine the model in which these
decays proceed through a gluonic component of the $J/\Psi$ and the
$\eta_c$, arising from a mixing of $(c\bar c)$ and glueball states.
We give some bounds on the values of the mixing angles and propose
the study of the $p \bar{p} \to \phi \phi$ reaction, at $\sqrt{s}
\simeq 3$ GeV, as an independent test of the model.

\vfill
\eject

\pagestyle{plain}

\vskip1cm
\section{Introduction}
\vskip1cm

It has been known for many years that the $J/\Psi$ and the $\eta_c$
decay respectively into Vector-Pseudoscalar (VP) and Vector-Vector (VV)
mesons, channels which should be forbidden by the helicity conservation
rule in perturbative QCD and the usual assumption of the dominance of
collinear valence quark configurations; a similar problem holds
for the decay of $\eta_c$ into $p \bar p$
\cite{bib:bl,bib:CZ1,bib:mauro1,bib:FH}.

Several attempts to overcome such difficulties by taking into account
and modeling non perturbative effects can be found in the literature
\cite{bib:mauro1,bib:diq,bib:mass}. Two quark correlations
inside the proton, or diquarks, although very likely present in the
$Q^2$ region of charmonium decays, do not seem to help with the
$\eta_c \to p\bar p$ large observed decay rate \cite{bib:diq}; quark
mass corrections are equally of little help with $\eta_c \to p\bar p$
\cite{bib:mass} and do not contribute at all to $\eta_c \to VV$ decays
\cite{bib:mauro1}.

Apart from eventual effects due to the intrinsic transverse motion of
quarks inside the hadrons, which have not yet been investigated, at
this stage two possible phenomenological explanations of the above
problems remain: one requires a large contribution due to twist three
operators or higher order Fock components of the hadron wave functions
\cite{bib:CZ1}, and the second assumes a mixing of the $J/\Psi$
\cite{bib:FH,bib:brod} and of the $\eta_c$ \cite{bib:trig} with a
glueball. We shall argue that, at the light of recent experimental data,
the latter assumption seems to us a more realistic explanation of the
puzzle; it certainly is a more interesting one, in that it amounts
to allow for the presence of the fundamental trigluonium states, which
have the quantum numbers $J^{PC} = 0^{-+} ,\ 1^{--},\ 3^{--}$, around
a mass of 3 GeV/$c^2$ \cite{bib:trig}, in agreement with the
prediction of some models \cite{bib:dec}.

In the next Sections, after a brief discussion of the higher twist
model and its comparison with data, we shall further investigate the
glueball mixing idea giving a first rough estimate of the values of
the mixing angles and proposing some experimental tests which could
shed some light on the subject in the near future.

\vskip0.3cm
\section{ Higher Twist Model}
\vskip1cm

In Ref. \cite{bib:CZ1} it has been suggested, in a consistent
perturbative QCD treatment, that the $J/\Psi$ decays into $VP$
could be explained through the introduction of twist three meson
wave functions, namely through the $\mid q \bar{q} g >$ Fock
components of the $V$ and $P$ mesons. Detailed calculations, in
good agreement with the data, have been performed for the
branching ratio $B(J/\Psi \to \rho\pi)$. Similarly, the double OZI
suppressed decay $\eta_c \to \omega \phi$ has been computed
and the result exploited, using $SU(3)$ symmetry and an effective
lagrangian, to estimate some relative magnitudes of other
$\eta_c \to VV$ branching ratios \cite{bib:CZ3}.

The first conceptual difficulty of this model is given by the necessity
of an unexpectedly large contribution from a higher Fock component,
also considering the theoretical problems connected with the study of
three-particle wave functions \cite{bib:CZ2}.
The main problem, however, is the fact that such scheme should
predict analogous results for the $\Psi'$ decays, whereas the decays
$\Psi' \rightarrow PV$ are observed to be strongly suppressed:
$Br(\Psi' \rightarrow \rho \pi) < 8.3 \cdot 10^{-5}$,
$Br(\Psi' \rightarrow K \bar K^{*}) <1.79 \cdot 10^{-5}$
\cite{bib:data}.

Furthermore, some other predictions of the model are not in agreement
with the experimental data:

\noindent
i) in Ref. \cite{bib:CZ3} the branching ratios of $\eta_{c}$,
$\chi_{0}$ and $\chi_{2}$ in the double OZI suppressed channel
$\omega \phi$ are predicted to be roughly of the same order of
magnitude as the single OZI suppressed ones. However, such double
OZI suppressed decays have not yet been observed, whereas the single
OZI suppressed decays into VV have been clearly observed.

\noindent
ii) Ref. \cite{bib:CZ3} predicts
\begin{equation}
{Br( \eta_c \rightarrow \omega \omega ) \over Br( \eta_c \rightarrow
\phi \phi) } \simeq 1.3
\label{eq:omomT}
\end{equation}
in disagreement with the experimental information
\begin{equation}
{Br( \eta_c \rightarrow \omega \omega) \over Br( \eta_c \rightarrow
\phi \phi) } < 0.44.
\label{eq:omom}
\end{equation}
Even inserting into Eqs. (38)-(39) of Ref. \cite{bib:CZ3} updated
experimental data, one obtains for Eq. (\ref{eq:omomT}) the
value $\simeq 0.8$, still in disagreement with Eq. (\ref{eq:omom}).

\noindent
iii) finally, no direct computation of $\eta_c \to VV \, (V,V
\not= \omega, \phi)$ with three-particle wave functions has been
attempted, and no attempt at all has been made to compute
the $\eta_c \to p \bar{p}$ decay rate.

We must accept that higher twist contributions still leave many
unsolved problems; we then turn to a different possible solution and
from now on we concentrate on the glueball mixing model.

\vskip0.3cm
\section{Mixing Parameters}
\vskip1cm

We adopt the following mixing schemes:

\begin{equation}
\mid J/\Psi > =  \cos\theta \mid c \bar c >_{_{J/\Psi}} +
\sin\theta \mid O>
\label{eq:mixi1}
\end{equation}

\begin{equation}
\mid \eta_c > =  \cos\alpha \mid c \bar{c} >_{_{\eta_c}}
+ \sin \alpha \mid P >
\label{eq:mixi2}
\end{equation}
where $O$ is a vector and $P$ a pseudoscalar glueball.

Due to perturbative QCD helicity conservation it follows:
\begin{eqnarray}
< VP \mid c \bar c >_{_{J/\Psi}} & = & 0 \\
< VP \mid J/\Psi > & = & \sin \theta < VP \mid O >
\end{eqnarray}
and due to the (lowest order) decoupling of gluons from photons,

\begin{eqnarray}
< e^{+} e^{-} \mid O > & = & 0 \\
< e^{+} e^{-} \mid J/\Psi > & = & \cos \theta < e^{+} e^{-}  \mid
c \bar c >_{_{J/\Psi}} \, .
\end{eqnarray}

Analogous equations are valid, {\it mutatis mutandis},
for the pseudoscalar case ($ O \to P, \> J/\Psi \to \eta_c, \>
VP \to VV, \> e^{+} e^{-} \to \gamma \gamma$).

In order to obtain an, admittedly rough, estimate of the mixing angles we
need at least an estimate of the total width of the $O$ ($P$)
and of the partial width of the $O$ ($P$) into Vector-Pseudoscalar
(Vector-Vector) mesons.
A possible way to get the former is offered by the $\sqrt{OZI}$
suppression of the glueball decay rule \cite{bib:sozi}, which suggests
the width of a trigluonium to be in between the one for a light quark
state with a mass around 3 GeV/$c^2$ and the $J/\Psi $ one,
namely \begin{equation}
\Gamma_O \approx \sqrt{ \Gamma_{J/\Psi} \cdot 500 \, {\rm MeV} }
 \approx 7 \, {\rm MeV} \label{eq:WT}
\end{equation}
and the width of a digluonium to be beteewn the one for a light quark
state with a mass around 3 GeV/$c^2$ and the $\eta_c $ one, namely
\begin{equation}
\Gamma \approx \sqrt{ \Gamma_{\eta_c} \cdot 500 \, {\rm MeV}}
 \approx 70 \, {\rm MeV} \label{eq:WD}
\end{equation}

Such a rule, however, has been criticized on various grounds
\cite{bib:sc}, and in the following we will assume the values in
Eq. (9) and (10) as lower bounds, taking
\begin{equation}
\Gamma_O \approx \Gamma_P \approx (10 \div 100) \, {\rm MeV}.
\label{eq:WOP}
\end{equation}
This estimate is consistent with the value suggested in Ref.
\cite{bib:brod}.

In order to estimate $\Gamma_{O \rightarrow VP}$ we consider
that a $1^{--}$ state is allowed to decay into eight different couples
of meson nonets:
$VP$, $SV$, $VT$, $PA^+$, $SA^-$, $VA^+$, $TA^-$, $A^+A^-$;
therefore, if we consider a full flavour democracy in the glueball's
decays and ignore differences in phase spaces, we have that
(approximating to 1/10 in order to include also eventual excited nonets
and three body decays)
\begin{equation}
\Gamma_{O \rightarrow VP} \approx {1 \over 10} \, \Gamma_{O} \approx
(1 \div 10) \, {\rm MeV}.
\label{eq:PWO}
\end{equation}

Analogously, for the pseudoscalar state we have 13 channels:
$PP$, $PS$, $PT$, $SS$, $ST$, $VV$, $TT$, $PA^-$, $SA^+$, $VA^-$,
$TA^+$, $A^+A^+$, $A^-A^-$,
and also in this case, within our approximations, we can assume
\begin{equation}
\Gamma_{P \rightarrow VV} \approx {1 \over 10} \, \Gamma_{P}
\approx (1 \div 10) \, {\rm MeV}.
\label{eq:PWP}
\end{equation}

Of course $\Gamma_{O \rightarrow \rho \pi}$ ($\Gamma_{P \rightarrow \rho
\rho}$) is a fraction of the above width; however, considered that the
decays into VP (VV) are favoured with respect to the decays into the
other couples of nonets by phase space, we can assume as a rough
estimate:

\begin{equation}
\Gamma_{O \rightarrow \rho \pi} \lsim (1 \div 10) \, {\rm MeV}.
\label{eq:PWrp}
\end{equation}

\begin{equation}
\Gamma_{P \rightarrow \rho \rho} \lsim (1 \div 10) \, {\rm MeV}.
\label{eq:PWPrr}
\end{equation}

Using Eq. (14) we obtain from Eqs. (3) and (6)
\begin{equation}
\Gamma_{J/\Psi \rightarrow \rho \pi} = 1.09 \, {\rm keV} \approx
\sin^2 \theta \cdot \Gamma_{O \rightarrow \rho \pi} \lsim
(1 \div 10) \cdot \sin^2 \theta \, {\rm MeV}
\label{eq:mix2}
\end{equation}
which leads to
\begin{equation}
\sin^2 \theta \gsim (10^{-4} \div 10^{-3})
\label{eq:stetg}
\end{equation}

On the other hand we must have
\begin{equation}
\Gamma_{J/\Psi} > \sin^2 \theta \, \Gamma_O
\label{eq:mix3}
\end{equation}
which gives

\begin{equation}
\sin^2 \theta < 8.6 \cdot (10^{-4} \div 10^{-3})
\label{eq:stetl}
\end{equation}

Analogously we have for the pseudoscalar case, from Eqs.
(\ref{eq:PWPrr}) and (\ref{eq:mixi2}):
\begin{equation}
\Gamma_{\eta_c \rightarrow \rho \rho} = 0.267 \, {\rm MeV} \approx
\sin^2 \alpha \cdot \Gamma_{P \rightarrow \rho \rho} \lsim
(1 \div 10) \cdot \sin^2 \alpha \, {\rm MeV}
\label{eq:mix2P}
\end{equation}
which leads to
\begin{equation}
\sin^2 \alpha \gsim 2.7 \cdot (10^{-2} \div 10^{-1}) \, ,
\label{eq:salpg}
\end{equation}
and
\begin{equation}
\Gamma_{\eta_c} > \sin^2 \alpha \, \Gamma_P
\label{eq:mix3P}
\end{equation}
which gives only the trivial bound

\begin{equation}
\sin^2 \alpha < (0.1 \div 1).
\label{eq:salpl}
\end{equation}



\vskip0.3cm
\section{Tests of the model}
\vskip1cm

The above results clearly show how a tiny admixture of glueballs in
$\eta_c$ and $J/\Psi$ could be sufficient to explain the otherwise
problematic charmonium decays. Let us then consider other consequences
and predictions of the mixing scheme.

It is clear that, away from the glueball mass region, the model under
investigation cannot help any more and perturbative QCD predictions
should be valid; as a consequence, one expects the decays $\eta_c'
\to VV, p \bar p$ \cite{bib:trig} and $\Psi' \to VP$ \cite{bib:brod} to
be quite suppressed, as the mass difference between these states and
the glueballs is much bigger and therefore the eventual mixing much
reduced. Indeed the branching ratios for the decays $\Psi' \to VP$
are known to be very small \cite{bib:data} and no $\eta_c'$ has been
observed up to now in $p \bar p$ annihilation \cite{bib:Cester}.

Of course, also the decays $\eta_b \rightarrow VV, p \bar p$ and
$\Upsilon \rightarrow VP$ should not be allowed, in agreement with
perturbative QCD, there being no reason to have a glueball in the
large mass range of these states.

A direct observation of the glueball state in the proximity
of the $\eta_c$ and $J/\Psi$ mass would certainly be decisive.
The best process where to look for a glueball in this
mass region is $p \bar{p} \to \phi \phi$, which is doubly OZI
forbidden and should proceed through a purely gluonic state.
The cross section for this reaction, if the glueball really appears as
an intermediate state in the s-channel, should manifest a resonant
behaviour.
In the case of the pseudoscalar glueball $P$, the cross section
at the peak can be written down as:
\begin{equation}
\sigma^{Peak}(p \bar{p} \to P \to \phi \phi)
= {\pi \over k^{2}_{CM}} Br(P \to \phi \phi) \, Br(P \to
p \bar{p})
\end{equation}
where $k_{CM}$ is the center of mass momentum of the initial hadrons.
In order to estimate, albeit roughly, this cross section we can use the
results of section 3 which give $Br (P \to \phi \phi ) \simeq 0.1$
and
\begin{equation}
\Gamma (P \to p \bar{p}) = {\Gamma (\eta_c \to p \bar{p})
\over \sin^{2} \alpha}.
\end{equation}
Using the experimental value of $Br (\eta_c \to p \bar{p})$
\cite{bib:data}  we get
\begin{equation}
\sigma^{Peak}(p \bar{p} \to P \to \phi \phi)
\simeq {1 \over \sin^2 \alpha} \, ( 10^{-2} \div 10^{-1} ) \,
\mu {\rm b}
\end{equation}
Remarkably, this value increases when $\alpha$ is small and therefore it
permits to establish even a small admixture of glueball in the
charmonium state.

Taking $\sin^2 \alpha \simeq 0.1$, according to the results of the
previous section, we obtain the numerical estimate
\begin{equation}
\sigma^{Peak}(p \bar{p} \to P \to \phi \phi) \simeq (10^{-1} \div
1) ~ \mu {\rm b}. \end{equation}

Some experimental data on the $p \bar{p} \to \phi \phi$ process
(although preliminary) are already available from the LEAR experiment
at CERN, up to a center of mass energy of 2.4 GeV \cite{bib:Macri}.
The value of the cross section is about 1 $\mu$b at the highest
energy available.
It is clearly seen from the data that this cross section decreases with
energy. Therefore, according to the our estimate of
$\sigma^{Peak}(p \bar{p} \to \phi \phi) $, the signal due to the
presence of a glueball resonance in the energy region of $\sqrt{s}
\simeq 3$ GeV should be observable above the
expected background.
Also the cross sections $\sigma(p \bar{p} \to J/\Psi \to \phi \phi) $
and $\sigma(p \bar{p} \to \eta_c \to \phi \phi) $ are much smaller than
the value estimated above.

The observation of this glueball state in the $ p \bar{p} \to \phi \phi$
reaction would therefore be a clear test of the model under
investigation.

For what concerns the vector glueball state, it should manifest itself,
for instance, in $\Psi' \to \rho \pi X$ decay as a resonance in the
$\rho \pi$ effective mass distribution.

Another possibility to observe gluonium is to look for a resonance
in the $\phi \phi$ effective mass distribution in the reaction
$\pi^{-} p \to \phi \phi X$.

Notice that the decays which can proceed through the $(c\bar c)$
component are not much affected
by the glueball mixing, since the mixing angle is very small,
as we have shown. We expect in these channels effects of the
order of $\approx 1 \% $, which, considering the
theoretical uncertaintes in the charmonium wave functions
\cite{bib:pot,bib:lat}, would be quite difficult to detect
even in the next generation experiments.

Finally it can be interesting to note that the mixing angles $\theta$
and $\alpha$ could be experimentally related through the ratio

\begin{equation}
{\Gamma(\eta_c \rightarrow \gamma \gamma) \over
\Gamma(J/\Psi \rightarrow e^+ e^-)} = 1/12 \cdot
{| \Psi_{\eta_c}(0)|^2 \over |\Psi_{J/\Psi}(0)|^2}
(1+1.96 {\alpha_s \over \pi}) {\cos^2 \alpha \over \cos^2 \theta}
\label{eq:at}
\end{equation}
which gives, assuming $|\Psi_{\eta_c}(0)| =
|\Psi_{J/\Psi}(0)|$,
\begin{equation}
{\cos^2 \alpha \over \cos^2 \theta}=0.78 \pm 0.29
\label{eq:at2}
\end{equation}
in agreement with a larger $\alpha$ than $\theta$.
Of course, this value could be much improved in the next future.
Incidentally, it must be noted that Ono and Sch\"orbel \cite{bib:OS}
claim different values for the ratio  $|\Psi_{\eta_c}(0)| /
|\Psi_{J/\Psi}(0)|$; however, their model seems to be excluded by
the recent data on the charmonium $^1 P _1$ mass \cite{bib:1p1}.

\vskip0.3cm
\section{Conclusions}
\vskip1cm

We can summarize our conclusions by stating that the only available
model which could explain the decays of $J/\Psi$ and $\eta_c$
forbidden in the usual perturbative QCD treatment of exclusive
processes is the one requiring the presence of gluonium states in the
proximity of their masses.

In this paper we have shown that such a model is not excluded by the
existing data; a tiny gluonic admixture can easily account
for the ``forbidden'' decays and does not significantly change the
allowed ones. The model can be easily
tested in that one should recover the usual perturbative QCD
predictions for the decays of large mass states: in particular
the helicity forbidden decays of $\eta_c'$, $\Psi'$, $\Upsilon$,
$\Upsilon(2S)$, etc. should be strongly suppressed.

A direct decisive test of the model can be provided by the experimental
observation of a (gluonic) resonance in the $p \bar{p} \to \phi \phi$
reaction at $\sqrt{s} \simeq 3$ GeV; we have given an estimate of
the corresponding cross section.

\end{document}